\documentclass[11pt,a4paper]{article}
\usepackage{epsf,feynmf,amsmath}
\usepackage{graphicx,color,mathrsfs,amsmath,amssymb}
\title{ \bf The fate of the trace anomaly in a finite formulation of field theory}
\author{Jean-Fran\c cois Mathiot \thanks{e-mail: jean-francois.mathiot@clermont.in2p3.fr} \vspace{0.3cm}\\
{\small \em  Universit\'e Clermont Auvergne, Laboratoire de Physique de Clermont,
} \\ {\small \em CNRS/IN2P3, BP10448, F-63000 Clermont-Ferrand, France}}
\date{}
\begin{document}
\maketitle
\bibliographystyle{unsrt}
\abstract{Within the framework of the recently proposed Taylor-Lagrange regularization scheme - which leads to finite elementary amplitudes in $4$-dimensional space-time with no additional dimensionful scales - we show that the trace of the energy-momentum tensor does not show any anomalous contribution even though quantum corrections are considered. Moreover, since the only renormalization we can think of within this scheme is a finite renormalization of the bare parameters to give the physical ones, the canonical dimension of quantum fields is also preserved by the use of this regularization scheme.}
%
\section{Introduction} \label{introduction}
The construction of the Lagrangian of the Standard Model as a quantum field theory relies on three main requirements: {\it i)} it should be {\it local}, {\it i.e.} it should be constructed from the product of fields or derivative of fields at the same space-time point; {\it ii)} it should respect at least two invariance principles: {\it Lorentz} and {\it gauge invariances}; {\it iii)} it should be {\it renormalizable}.
The success  of the Standard Model in the present understanding of  experimental results in a wide range of energy scales is a first indication that these three requirements may also be at the heart of our understanding of the microscopic laws of Nature. 

From a theoretical point of view, the accurate calculation of physical observables - as required for instance by precision experiments - does impose a careful understanding, and control, of loop corrections.  These ones are known - from the very beginning of $QED$ \cite{schwi} - to generate divergencies associated, for instance, to  infinitely large internal momenta. Surprisingly enough, the way to deal in practice with these divergencies has not much changed since the early days of $QED$. Although an elegant way to circumvent these divergencies has been proposed in the 70's with the advent of dimensional regularization ($DR$) \cite{DR_1,DR_2,DR_3,DR_4}, one is still discussing in the literature radiative corrections in terms of a na\"ive cut-off in momentum space, or by adding non-physical degrees of freedom like Pauli-Villars ($PV$) fields.

This is a clear indication that the very first origin of these divergencies has not been properly taken care of. In all the above cited regularization procedures infinities are indeed recovered in physical conditions \footnote{These physical conditions imply working in a 4-dimensional space-time, with infinite cut-off or infinite $PV$ masses.} for any elementary amplitude . The origin of these divergencies is however well known since the 50's \cite{bogol,bosh,stuec}. Any quantum field should be considered as an operator valued distribution ($OPVD$) \cite{wigh,haag,schw}, so that the construction of a local Lagrangian should be done within the general framework of the theory of distributions \cite{schwa} (previously known as generalized functions \cite{sobol}). Otherwise, the product of two fields at the same point is ill-defined from a mathematical point of view, leading to the observed divergencies in loop calculations as well as violation of causality \cite{tkach,aste}.

The taming of these  divergencies involves a so-called regularization procedure. This procedure should preserve, to start with, the symmetry principles of the Lagrangian. Using a na\"ive cut-off for instance is known to violate Lorentz and gauge invariances, while using $DR$ does preserve these fundamental symmetries. This different behavior should be related to the clear distinction between these two regularization procedures: while $DR$ is implemented ``a-priori'' in the construction of the Lagrangian density at $D=4-\epsilon$ dimensions, the use of a na\"ive cut-off is implemented ``a-posteriori'' at the level of the calculation of each individual (divergent) amplitude. The first regularization procedure is thus called an {\it a-priori}  regularization procedure, while the second one is called {\it a-posteriori}  \cite{axial}.
The use of $DR$ does not however address directly the origin of these divergencies. As mentioned above, it just avoids to treat the problem by ``escaping'' in $D=4-\epsilon$ dimensions! While physical observables are finite at $D=4$ after mass, coupling constant and field renormalization - thanks to the renormalizability of the Standard Model - individual, elementary, amplitudes are still divergent in physical $D=4$ conditions. 

The way to properly implement the nature of quantum fields as $OPVD$ in practical calculations has only been developed recently \cite{EG, schar, GB, GW_mass,GW}. We shall concentrate in this study in the Taylor-Lagrange regularization scheme ($TLRS$) proposed in Refs.~\cite{GW_mass} and \cite{GW}. The  application of this scheme to the calculation of the radiative corrections to the Higgs mass \cite{FT}, or to the recovery of the axial anomaly \cite{axial}, is a clear illustration of how this regularization scheme should be used in practice. This procedure is called ``{\it intrinsic}'' since each individual, elementary, amplitude is finite in physical, $D=4$, conditions, hence the denomination of finite formulation of field theory. By contrast, $DR$ is called ``{\it extrinsic}'' since the divergencies are only taken care of, at $D=4$, for  the full physical observable - after mass, coupling constant and wave-function renormalization - and not at the level of each individual amplitude. Note that thanks to the above properties, $TLRS$ is at the same time a regularization procedure as well as a renormalization scheme, with the same acronym. The only renormalization constant we can think of within this scheme is a purely finite renormalization which relates bare parameters to physical ones, as in any many-body interacting systems. As explained in  Ref.~\cite{jfm}, the (finite) bare coupling constant in this scheme does depend on a {\it dimensionless} scaling variable called $\eta$. This is at variance with the physical coupling constant which depends on the (dimensionful) energy scale at which it is measured. \\

Apart from inducing divergencies of elementary amplitudes, quantum corrections are known to generate anomalies: an anomaly is said to occur when the conservation of a given current at tree level is violated by quantum corrections. Two anomalies have been identified in quantum field theory for instance: the axial anomaly and the trace anomaly. 
The axial anomaly appears in an axial-vector-vector transition. At tree level, and for massless particles, the axial current associated to this transition is conserved, as well as the vector current. Beyond tree level,  the conservation of the axial current is violated, while the vector current should be conserved, thanks to gauge invariance. Such behavior is indeed confirmed by experiment, from the $\pi_0 \to \gamma \gamma$ decay amplitude. The calculation of such anomaly for the axial current, together with the conservation of the vector current, are therefore  a strong test of any regularization procedure. This check is known since a long time in the case of $DR$ \cite{adle}. It has also been verified  very recently when $TLRS$ is implemented \cite{axial}.

We shall concentrate in this study on the second anomaly, the so-called trace anomaly. Contrarily to the axial anomaly, this one is not confirmed by an experimental evidence. Its status should therefore be  discussed in the light of the main properties of each regularization procedure. In the  limit of massless particles,  the dilatation current associated to the scaling of space-time coordinates is conserved at tree level (see for instance Ref.~\cite{cole}). This conservation is violated when various regularization procedures are considered, like for instance using  $PV$ regularization \cite{call}, or $DR$ \cite{adle}. {\it We shall show in this study that this is indeed not the case when $TLRS$ is used}. This is a direct consequence of the unique properties of $TLRS$: it does not involve any new dimensionful parameter, and it can be implemented directly in physical $D=4$ space-time dimensions. This is a unique feature of this {\it a-priori} as well as {\it intrinsic} regularization procedure.
From these general properties, we   show that the canonical dimension of the quantum fields is preserved by quantum corrections, in contrast also to the common belief. 

The plan of our article is the following. We show in Sec.~\ref{trace} that the trace anomaly is just an artefact of the regularization procedure and disappears when  $TLRS$ is used. We treat in Sec.~\ref{dimension} the case of the effective dimension of  quantum fields. Our conclusions are drawn in Sec.~\ref{conc}.  

\section{The trace anomaly} \label{trace}
The trace anomaly is associated to the non-conservation of the dilatation current, in the limit of massless particles, when quantum corrections are considered. In order to calculate these quantum corrections, as mentioned in the Introduction, one should first choose an appropriate regularization procedure in order to give a well defined mathematical meaning to the local Lagrangian we start from. We have organized these (a-priori) procedures in two different classes: the {\it extrinsic} regularization procedures, like $PV$ or $DR$, for which divergencies are recovered for each elementary amplitude in physical conditions, and the {\it intrinsic} ones, like $TLRS$, where each elementary amplitude is finite in four space-time dimensions with physical degrees of freedom only. As we shall see below, these two classes behave quite differently as far as the trace anomaly is concerned.

As detailed in Ref.~\cite{GW}, the physical field $\varphi$ is constructed in $TLRS$ as a functional of the original quantum field $\phi(x)$ - considered as a distribution - according to
\begin{equation} \label{distri}
\varphi[\rho](x)\equiv \int d^4 y \phi(y) \rho(\vert x-y \vert),
\end{equation}
where the reflection symmetric test function $\rho$ belongs to the Schwartz space  $\mathscr{S}$ of rapidly decreasing functions \cite{schwa}. We consider here for simplicity a scalar field $\phi$, the construction of gauge fields being detailed in Ref.~\cite{GW_gauge}. The physical interest to use the test function $\rho$ is to smear out the original distribution in a  space-time domain of typical extension $a$. The test function can thus be characterized by $\rho_a(x)$ and the physical field is simply written $\varphi_a(x) \equiv \varphi[\rho_a](x)$. In order to recover the local character of the original Lagrangian, we shall take - at the level of each individual amplitude - the {\it continuum limit} $a\to 0$. The only remnant of such limit is a {\it dimensionless} scaling variable $\eta$ since  we also get $a/\eta \to 0$ for any $\eta > 1$. We thus have in this limit $\varphi_a \to \varphi_\eta$.  For simplicity, we shall remove the subscript $\eta$ in the following. All bare parameters of the Lagrangian like masses and coupling constants do depend on this scaling variable, called also regularization scale. 

Under a scale transformation of the space-time coordinate $x^\mu$ defined by 
\begin{equation} \label{scalei}
x^\mu \to e^\alpha x^\mu,
\end{equation}
the quantum field $\phi(x)$ should transform according to
\begin{equation} \label{sca}
\phi(x) \to e^{\alpha d_\phi^0} \phi(e^\alpha x),
\end{equation}
where $d_\phi^0$ is the canonical dimension of the bare field $\phi$. The  field $\varphi$ transforms  according to
\begin{eqnarray}
\varphi(x) \to && e^{\alpha d_\phi^0}\int d^4 (e^\alpha y) \phi(e^\alpha y) \rho(e^\alpha \vert x-y \vert)\nonumber \\
&&=e^{\alpha d_\phi^0}\int d^4 z \phi(z) \rho(\vert e^\alpha x-z \vert) \equiv e^{\alpha d_\phi^0} \varphi(e^\alpha x).
\end{eqnarray}
The construction of the  field $\varphi$ in (\ref{distri})  respects also the scaling transformation (\ref{sca}).
Under an infinitesimal scaling transformation, we thus have \cite{cole}
\begin{equation}
\varphi(x) \to \varphi(x) + \alpha(d_\phi^0+x^\mu \frac{\partial}{\partial x^\mu})\varphi(x).
\end{equation}
Note that the above equation, with its derivative on $x^\mu$, is only defined with well behaved functionals $\varphi$. This is indeed the case in $TLRS$ with the construction of the physical field in Eq.~(\ref{distri}). This is however not the case when using the original quantum field $\phi(x)$, defined as an $OPVD$,  as it is usually done.

The current associated to this transformation is given by
\begin{equation} \label{dila}
D^\mu=x_\nu \Theta^{\mu \nu},
\end{equation}
where $\Theta^{\mu \nu}$ is the symmetric energy-momentum tensor \cite{call,call_2,cole_2}.  
The divergence of this current is thus simply given at tree level by
\begin{equation} \label{traced}
\partial_\mu D^\mu=\Theta_\mu^{\ \mu}=m_0^2 \varphi^2 \equiv \Delta,
\end{equation}
where $m_0$ is the bare mass of the scalar degree of freedom. 
The extension of (\ref{traced}) beyond tree level is at the origin of the so-called trace anomaly. 

The calculation of the dilatation current (\ref{dila}) and its divergence (\ref{traced}) implies to start from a well defined Lagrangian. This is done by choosing an {\it a-priori} regularization procedure like using $PV$ fields, $DR$, or $TLRS$. 
As an {\it intrinsic} regularization procedure however, $TLRS$ has unique properties as compared to $DR$ or $PV$ regularization.
\begin{itemize}
\item Since $TLRS$ is at the same time a regularization procedure as well as a renormalization scheme, the calculation of the dilatation current, and its derivative, can be done at the level of the bare fields and not at the level of the renormalized ones. Both $PV$ regularization and $DR$  require an (infinite) wave-function renormalization. This implies to work with renormalized fields in order to manipulate well defined mathematical quantities in physical conditions. This is not the case when using $TLRS$ since all bare quantities are finite in this scheme.

\item $TLRS$ operates in physical $D=4$ dimensions with no additional non-physical mass scales. We cannot therefore generate any new contribution to the derivative of the dilatation current, as it is the case for instance in $DR$ or $PV$ regularization.
\item In the continuum limit, each elementary amplitude is finite in $TLRS$, and exhibits a dimensionless scaling variable $\eta$. All the bare parameters and fields depend on this arbitrary dimensionless regularization scale. As explained in Ref.~\cite{jfm},  the physical coupling constant only should depend on a dimensionful variable, like the (physical) renormalization point in the on-mass shell renormalization scheme for instance.
\end{itemize}
From the above general properties of $TLRS$, we can immediately derive the equations relating the ($1PI$) Green's functions $\Gamma^{(n)}$ and $\Gamma_\Delta^{(n)}$, where $\Gamma_\Delta^{(n)}$ corresponds as usual to the $1PI$ Green's function associated to the insertion of the composite operator $\Delta$. From general dimensional arguments on the one hand, for a field of mass $m_0$, we can write, using $TLRS$ at $D=4$,
\begin{equation} \label{scale}
\Gamma^{(n)}(k_1 \ldots k_n)=m_0^{4-n d_\phi^0}f^{(n)}\left[\frac{k_1}{m_0}, \ldots , \frac{k_n}{m_0}\right],
\end{equation}
where $f^{(n)}$ is a dimensionless function. Since we operate at $D=4$ dimensions, the bare coupling constant is dimensionless, and there is no other dimensionful scale apart from the bare mass $m_0$. One cannot  therefore generate any new contribution to the equation of dimensional analysis. As well known, this is in contrast with the use of $DR$ where the coupling constant acquires a dimension $\mu^\epsilon$ where $\mu$ is a new arbitrary mass unit \cite{thoof}, or with the use of $PV$ fields which involves their (infinitely large) masses. We can thus deduce the following equation of dimensional analysis
\begin{equation} \label{dimen}
\left[\sum_{i=1}^{n}k_i. \frac{\partial}{\partial k_i}\right] \Gamma^{(n)} (k_1\ldots k_n)=(4-n d_\phi^0) \Gamma^{(n)}(k_1\ldots k_n)-m_0\frac{\partial}{\partial m_0}\Gamma^{(n)}(k_1\ldots k_n).
\end{equation}
On the other hand, from the trivial identity \cite{cheng} between bare operators
\begin{equation}
m_0\frac{\partial}{\partial m_0} \Gamma^{(n)}(k_1 \ldots k_n)=-i\Gamma_\Delta^{(n)}(0;k_1\ldots k_n),
\end{equation}
we can  write, with Eq.(\ref{dimen}),
\begin{equation} \label{vertex}
\left[\sum_{i=1}^{n}k_i. \frac{\partial}{\partial k_i}-(4-nd_\phi^0)\right]\Gamma^{(n)}(k_1\ldots k_n)=i\Gamma_\Delta^{(n)}(0;k_1\ldots k_n).
\end{equation}
We recover as expected the Ward identity \cite{call_2} associated to (\ref{dila}), with the canonical dimension $d_\phi^0$ (see next Section).
As a direct consequence of the above properties, there are no new contributions to the divergence of the dilatation current, apart from the trivial mass term. The conservation of the dilatation current {\it is thus preserved} in the massless limit when quantum corrections are calculated within the general framework of $TLRS$. This is at variance with the common belief, using for instance $DR$ or  $PV$ regularization procedures. This finding is a direct consequence of the unique properties of $TLRS$. As an {\it a priori} regularization procedure, it enables to start from a  (mathematically) well defined local Lagrangian, and as an {\it intrinsic} one it leads to finite bare elementary amplitudes in physical conditions, without  introducing any new arbitrary  mass scales. This is indeed what is expected from general arguments since, starting from a (well defined) local Lagrangian without any new mass scale apart from the bare mass of the physical particles, the ultra-violet ($k \to \infty$) as well as infra-red ($k\to 0$) limits are both invariant under the dimensionless scaling transformation $k\to \eta k$.

\section{Dimension of quantum fields} \label{dimension}
These unique properties of $TLRS$ have also immediate consequences as far as the dimensionality of the quantum fields is concerned. It is usually accepted that quantum corrections do generate an anomalous dimension of the quantum fields on top of its canonical dimension $d_\phi^0$\cite{wils}. In the common treatment of divergencies, using $DR$ for instance, this anomalous dimension is associated to the necessary (infinite) renormalization of the fields in order to be able to remove all divergencies in the calculation of physical observables in the limit  $D\to4$. 

The situation is quite different when $TLRS$ is used. Starting from a well defined bare Lagrangian at $D=4$ - with quantum fields having their canonical dimension - any elementary amplitude is finite. The only renormalization of the fields we can therefore think of is a finite renormalization constant when the on-mass shell renormalization scheme if used for instance. {\it This finite renormalization cannot lead whatsoever to a correction to the effective dimension of any quantum field}. This is in complete agreement with the above derivation of Eq.~(\ref{vertex}).

As already mentioned in Ref.~\cite{jfm}, the renormalization group ($RG$) equation associated to the regularization scale $\eta$  concerns the $S$ matrix only, since the bare $n$-point $1PI$ Green's functions  depend on this scale through the $\eta$-dependence of the bare parameters.  This $RG$ equation does not involve any correction associated to the coefficient called   ``anomalous dimension'' in the literature and defined in $TLRS$ by
\begin{equation} \label{and}
\gamma_\phi=\frac{\eta}{2Z_\phi} \frac{\partial Z_\phi}{\partial \eta},
\end{equation}
where $Z_\phi$ is the finite field renormalization constant in $TLRS$.
This coefficient $\gamma_\phi$ is  non zero, and finite, as calculated for instance  in $QED$ in first order perturbation theory \cite{GW_gauge}. This coefficient is indeed mandatory in order to get $\eta$-independent physical observables as it should. We  emphasize that although physical fields keep their canonical dimensions when quantum corrections are calculated within $TLRS$, the coefficient called ``anomalous dimension '' and defined in (\ref{and}) is non zero, but finite.

\section{Conclusions} \label{conc}
The use of $TLRS$ in order to calculate quantum corrections shed a completely new light on our understanding of the dimensionality of quantum fields as well as on the interpretation of the trace anomaly. This is a direct consequence of the unique properties of $TLRS$ as a finite formulation of field theory. As an {\it a priori} regularization procedure, it is implemented at the level of the original Lagrangian we start from and not at the level of each individual amplitude, and as an $\it intrinsic$ one, it leads to finite, elementary, amplitudes in physical $D=4$ dimensions. These amplitudes do depend on a dimensionless scaling variable $\eta$. This is in contrast with all other regularization procedures. These  unique properties - finiteness of any individual amplitude in physical $D=4$ dimensions together with no additional dimensionful scale - imply quite naturally that the physical fields do conserve their canonical dimension while the trace of the energy-momentum tensor is conserved in the limit of massless particles even though quantum corrections are considered, hence the absence of the so-called trace anomaly.

It is interesting at this point to compare our results with the ones using $DR$ in the minimal subtraction scheme ($MS$). From a practical point of view, for the calculation of a typical physical observable in perturbation theory for instance, one may argue that the use of $TLRS$ or $DR+MS$  provides very similar results. The reason is simple: while in $TLRS$ the divergent contributions are absent by construction, these contributions are  cancelled {\it by hand} with the use of appropriate counterterms in $DR+MS$. This similarity is indeed true as long as no multiplicative cancellations  occur between poles in $1/\epsilon$ and terms proportional to $\epsilon$ in $DR$ in the operators we are dealing with. In this case, however, one may get a spurious constant contribution which can be interpreted as physical. This is  what happens in the standard calculation of the trace anomaly in $DR+MS$ \cite{adle}. The use of $TLRS$ in  $D=4$ conditions insures that non physical contributions do not contribute at all. 

As already noticed in Ref.~\cite{adle}, axial and trace anomalies refer to very different situations. This explains why the axial anomaly is preserved in $TLRS$ - in accordance with experimental results - while the trace anomaly is not. The axial anomaly is associated to the calculation of a physical amplitude (the so-called triangle diagram) which exhibits a non trivial and constant dimensionless contribution in the ultra-violet domain. This contribution is associated, in $TLRS$, to the asymptotic properties of the test functions \cite{axial}.  The trace anomaly on the other hand, is associated, when $DR$ is used for instance, to spurious contributions appearing already at the level of the divergence of the dilatation current, as calculated from the local Lagrangian we start from. These contributions are absent, by construction, in $TLRS$.
Note that in the case of the axial anomaly the physical contribution to the $\pi \to \gamma \gamma$ amplitude is recovered directly in four dimensions in $TLRS$ \cite{axial} while it necessitates an adequate definition of $\gamma_5$ when $DR$ is used \cite{collins}.\\

From the due account of the nature of quantum fields as  $OPVD$, the use of $TLRS$ enables a direct calculation of physical observables in physical four space-time dimensions which are safe from ultra-violet as well as infra-red singularities. This general framework leads to a completely new understanding of various properties of quantum field theory beyond tree level. We show in this study that it leads to the absence of the so-called trace anomaly, while recovering the canonical dimension of the fields. This is made possible through the dimensionless nature of the regularization scale $\eta$, leading to a non-zero coefficient $\gamma_\phi$ called ``anomalous dimension'' in the standard formulation of the $RG$ equation, but this coefficient does not contribute at all to the effective dimension of the physical fields when quantum corrections are considered in $TLRS$.

\section* {Acknowledgement}
Preprint of an article published in Int. J. Mod. Phys. A {\bf 11} (2021) 2150265  (https://doi.org/10.1142/S0217751X21502651) © World Scientific Publishing Company.


\end{document}